\documentclass[prl,floatfix,twocolumn,showpacs]{revtex4}

\usepackage{graphicx}%
\usepackage{dcolumn}
\usepackage{amsmath}
\begin{document}

\preprint{cond-mat/none so far}

\title[Short Title]{Controlling spatiotemporal chaos in excitable media 
using an array of control points}
\author{S. Sridhar}
\author{Sitabhra Sinha}
\affiliation{%
The Institute of Mathematical Sciences, C.I.T. Campus, Taramani, Chennai - 600 113, India
}%

\date{\today}
\begin{abstract}
The dynamics of activation waves in excitable media can give rise to 
spiral turbulence, the resulting spatiotemporal chaos being
associated with empirical biological 
phenomena such as life-threatening disturbances in the natural rhythm of
the heart. 
In this paper, we propose a 
spatially extended but
non-global scheme using an array of control points 
for terminating such spatiotemporally chaotic excitations. 
A low-amplitude control signal is applied sequentially at each point on
the array, resulting in a traveling wave of excitation in the
underlying medium which drives away the turbulent activity.
Our method is robust even in the presence of significant heterogeneities
in the medium, which have often been an impediment to the success of
other control schemes.
\end{abstract}
\maketitle

Excitable media are a class of models for a wide range of physical,
chemical and biological systems that show spontaneous formation
of spatial patterns, such as spiral waves \cite{Keener}. These
spiral waves, under certain conditions, may become unstable
and break up, giving rise to spatiotemporal chaos. Such phenomena have been
seen to occur in a large variety of natural systems described
by excitable media, e.g., through 
interactions between
chemical concentration waves in the Belusov-Zhabotinsky 
reaction~\cite{Zaik70} and CO oxidation on Pt(110) surface~\cite{Jaku90},
cyclic AMP waves involved in the organization of multicellular
morphogenesis in {\em Dictyostelium}~\cite{Sieg95},
intracellular calcium waves in {\em Xenopus} oocytes~\cite{Lech91},
intercellular calcium waves in brain slices~\cite{Harr98,Huan04}, and
propagating electrical activity in the pregnant uterus~\cite{Lamm97}
and cardiac tissue~\cite{Gray98,Witk98}. The last example is of particular
importance as spatiotemporally varying patterns of excitation 
have been implicated in clinically significant disturbances of 
the natural rhythm of the heart, which are termed as arrhythmias. 
In fibrillation, a potentially life-threatening arrhythmia,
there is
complete loss of coordination between different regions of the heart
due to spatially extended chaotic activity \cite{Winfree}. 
The resulting cessation in the
mechanical pumping action necessary for blood circulation,
can lead to death within minutes unless
the condition is terminated promptly. 
Conventional defibrillation is done by applying 
a large electrical shock across the heart, which
is not only extremely painful, but can also damage cardiac tissue. 
Thus, devising a low-amplitude control method for
spatiotemporal chaos in excitable media is an exciting challenge,
as well as, having potential clinical 
relevance~\cite{Christini01,Gauthier02,Pumir07}.

Low-amplitude control of chaos in excitable media needs to take into
account certain special features of such systems, which hinder the
straightforward application of methods successful in other chaotic
systems. In particular, there exists a threshold for the stimulation,
only on exceeding which excitation occurs as indicated by a characteristic 
{\em action potential}. 
Thus, the control signal amplitude needs to be above a lower bound
to effect discernible changes in the medium, ruling out
simple proportional feedback schemes. The other complicating factor
is the existence of a {\em recovery period} immediately following excitation, 
during which the medium will
not respond to a normal supra-threshold stimulus. This implies that two
excitation wavefronts will annihilate on collision, as neither can
penetrate the other, their wavebacks being in the recovery phase.
This recovery period is distinguished
into {\em absolute}, during which the cell is insensitive to any perturbation, 
and
{\em relative}, when a larger than normal stimulus can excite the cell,
the amplitude depending on the exact phase of recovery.
Therefore, the amplitude and timing of the control signal needs to
be appropriately chosen, so that it results in a response from the medium.
Note that, the excited state is metastable, and the cell eventually recovers
to the {\em resting} state associated in different biological systems
with a characteristic resting transmembrane potential ($\simeq -84$ mV for
cardiac myocyte cells).
Thus, the control of spatiotemporal chaos in excitable media is essentially
a problem of synchronizing the excitation phase of every cell, so that the 
entire system returns to the resting state, resulting in the
termination of all activity. 

Chaos control schemes for excitable media
may be broadly classified into
\cite{Sinha06}: (i) {\em global}, where the control signal 
is applied to all points of the system, (ii) {\em local}, where only a small 
localized region of the system is subject to control, and (iii) 
{\em spatially extended but non-global} schemes.
Non-global methods use less power and also are relatively easier to implement 
practically, needing fewer control points.
However, strictly local control methods
almost always involve very high-frequency stimulation \cite{Zhang05}, that
can by itself lead to reentrant waves in the presence of 
inhomogeneities~\cite{PanfK93,Shajahan}. 
Moreover, the effect of local stimulation at a point can affect the rest 
of the system only through diffusion. 
As wavefronts annihilate on collision,
control-induced waves are restricted to the local neighborhood
of the stimulation point during spiral turbulence, with
the existing excited fragments closer to the control point shielding 
chaotic activity further away.
By using a spatially extended but non-global
scheme \cite{Sinha01} one can potentially avoid these drawbacks.
In this paper, we terminate chaos in excitable media by
applying spatiotemporally varying stimulation along an array of control
points. The control signal appears to propagate along the array, triggering
an excitation wavefront in the underlying medium, that is
regenerated after each collision with chaotic fragments and in the process 
eliminating all existing activity.
Stimulating each point once (or at most, twice) is seen to successfully
control chaos in almost all instances.
Although an array of control points have been used earlier to prevent the
breakup of a single spiral \cite{Rappel99}, to the best of our knowledge 
this is the first instance showing control of fully developed spatiotemporal 
chaos in excitable media using only a finite number of control points
without repeated stimulations at high frequency.

The spatiotemporal dynamics of excitation in several biological
systems can be
described by:
\begin{equation}
\partial{V}/\partial{t}= \frac{-I_{ion} + I_{ext} (x,y,t)}{C_m} + D \nabla^2 V,
\end{equation}
where $V$ (mV) is the transmembrane potential, $C_m$ = 1 $\mu$F cm$^{-2}$ 
is the
transmembrane capacitance, $D$ (cm$^2$s$^{-1}$) is the diffusion constant, 
$I_{ion}$ ($\mu$A cm$^{-2}$) is the transmembrane ionic current density
and $I_{ext} (x,y,t)$ is the space- and time-dependent external stimulus
current density applied for control on a 2-dimensional surface. 
For the specific functional form of 
$I_{ion}$, we used
the Luo-Rudy I (LR1) action potential model~\cite{Luo}, where, it is assumed
to be composed of six distinct currents,
each of them 
being determined by several time-dependent ion-channel gating variables $\xi$
whose time-evolution is described by differential equations $\frac{d\xi}{dt}
= \frac{\xi_{\infty} - \xi}{\tau_{\xi}}$. The parameters in these equations
are the steady-state values of $\xi$, 
$\xi_{\infty} = \alpha_{\xi}/(\alpha_{\xi} + \beta_{\xi})$, and the time
constants, $\tau_{\xi} = 1/(\alpha_{\xi} + \beta_{\xi})$, which are
governed by the voltage-dependent rate constants for the opening
and closing of the channels, $\alpha_{\xi}$ and $\beta_{\xi}$,
themselves complicated functions of $V$.
In order to verify the model independence of our results and to carry out
three-dimensional simulations, we have also used 
a simpler description of the action potential, as given 
by Panfilov (PV)~\cite{Panfilov93,Panfilov98}: $I_{ion} = f (V) - g$, where
$f$ is a piecewise linear approximation of a cubic function
and $g$ is an effective membrane conductance evolving with time as
$\frac{dg}{dt} = \epsilon(V,g) (kV-g)$. The
time-constant $\epsilon$ is a function of both $V$ and $g$ and
the parameters used are same as that in Ref.~\cite{Sinha01}. 
The models are solved using a forward-Euler scheme, the system being
discretized on a spatial grid with spacing $\delta x$ (=0.0225 cm for LR1,
= 0.05 cm for PV). The simulation domain is
a square lattice of $L \times L$ points in two dimensions or a cuboid with
$L \times L \times L_z$ points in three dimensions. For the LR1 simulations,
$L$ = 400, while for PV, $L$ = 256 (for 2-d) and $L=128, L_z = 8$ (for 3-d).
The standard five-point and seven-point difference stencils are used for the 
Laplacian in two and three dimensions, respectively. The time step 
for integration is chosen to be $\delta t$ = 0.01 ms (for LR1) and = 0.11 ms
(for PV).
No-flux boundary conditions are implemented at the edges of the simulation
domain. 
The initial spatiotemporally chaotic state is obtained 
by creating a broken wavefront which evolves into a spiral wave and 
is then allowed to become unstable, eventually breaking up into multiple
wavelets (in LR1 this process takes, on average, 200 ms).

\begin{figure}
  \includegraphics[width=0.95\linewidth, clip]{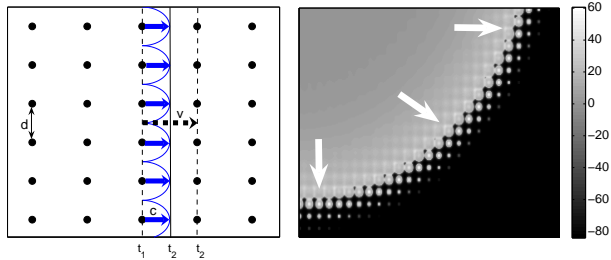}
  \caption{(left) Schematic diagram and (right) pseudo-gray-scale plot of 
  the transmembrane potential $V$ for the two-dimensional LR1 model, showing
  the propagation of the control-induced excitation wavefront. The 
  control signal traveling with velocity $v$ stimulates at time $t_1$
  a column of points (spaced $d$ apart) from which secondary excitations 
  traveling with velocity $c$ are generated. At time $t_2$
  the control signal stimulates the next column of points, which the induced
  excitation wave may or may not have reached depending on the relative
  values of $v$ and $c$ [$v/c = 4$ in (right)].
  }
\label{huyghens}
\end{figure}

We now focus on the control term $I_{ext} (x,y,t)$. For a 2-d domain of
size $L \times L$ we consider 
\begin{equation}
I_{ext} = I(x,y,t) \delta(x-md) \delta(y-nd),
\end{equation}
where, the delta function is defined as 
$\delta (z) = 1$ if $z=0$, and $=0$, otherwise, $d$ is the
spatial interval between points in an array where the control signal is 
applied and
$m,n$ are integers in the interval $[0,L/d]$. The current density
$I(x,y,t) = I_0$ for $t \in [\sqrt{x^2+y^2}/v,(\sqrt{x^2+y^2}/v)+\tau]$, 
and $= 0$ otherwise, corresponds to a rectangular control pulse of 
amplitude $I_0$ of duration $\tau$
that is travelling with velocity $v$. 
At the onset of control ($t = 0$), the point at $(0,0)$ is stimulated,
followed a short duration later by the points at $(0,d)$ and $(d,0)$,
and this process continues as the control pulse proceeds like 
a traveling wave across the array. 
At each control point, the stimulation may excite the underlying region 
depending on its recovery phase. An excited region can
in turn spread the effect of the stimulation to the
surrounding regions through diffusion. 
Fig.~\ref{huyghens} shows this process of secondary wave generation
at the stimulated control points, creating
a sustained excitation wavefront. 
Note that,
any portion of this stimulated wavefront in the medium
that is broken through collision with chaotic fragments, can be 
regenerated by
subsequent control points. Hence, there is effectively an unbroken
wavefront that travels through the medium, sweeping away the spatiotemporal
chaos and leaving the system in a recovering state.
As each control point needs to impose order over a
region of size $\sim d^2$ to eliminate spatiotemporal chaos, this highlights
the critical role of $d$. Indeed, for large $v$,
the method approaches global control as $d \rightarrow 0$, while for 
$d \sim L$ the control stimulation is confined to a single, localized region.
This implies that as $d$ increases, terminating chaos becomes increasingly
difficult. For example, in LR1 model, chaos control fails for $d \geq 13$.

Before reporting the simulation results, we consider the role played by the
traveling wave nature of the control pulse. The success of the proposed method
depends on the control signal velocity, $v$, relative to the velocity 
of the excitation wavefront
propagating via diffusion, $c$ (= 60.75 cm s$^{-1}$ for LR-1, 
= 50.9 cm s$^{-1}$ for PV, for
the parameters used here). When $v \rightarrow 0$, the control method
reduces to a local scheme regardless of $d$, as the effect of the external
stimulation can only propagate in the system via diffusion. On the other
hand, when $v$ is very large, all the control points are stimulated 
almost simultaneously. 
While the traveling wave nature of the control pulse allows
propagation of stimulation independent of diffusion through the excitable 
medium, for $v \simeq c$ the excitation propagating by diffusion reinforces
the external stimulation at each control point. 
Further, 
for a control signal propagating with a finite velocity (thus 
engaging only a few points at any given time), 
the energy applied per
unit time to the medium is much lower than that for simultaneous stimulation
of all the control points (i.e., $v \rightarrow \infty$).

For a system undergoing chaotic activity, the medium will at any time be at an
extremely heterogeneous state, with certain regions excited and other
regions partially or fully recovered. 
The vital condition for successful termination of chaos is that after the
passage of the control-stimulated wave there
should not remain any unexcited region which is 
partially recovered and
can be subsequently activated by diffusion from a decaying excitation front.
This places a lower bound on the control signal parameters, 
i.e., the signal amplitude $I_0$ and its duration $\tau$. If either is decreased
below this bound, the external stimulation is unable to excite certain
partially recovered regions. If these regions have neighboring
chaotic fragments, whose activity is slowly decaying after collision with
the control-induced wave, then, there will be a diffusion current
from the latter. Depending on the phase of recovery, this may be sufficient
to stimulate activity in the partially recovered regions,
thereby re-initiating spatiotemporal chaos after the control signal has
passed through.
\begin{figure}
  \includegraphics[width=0.95\linewidth, clip]{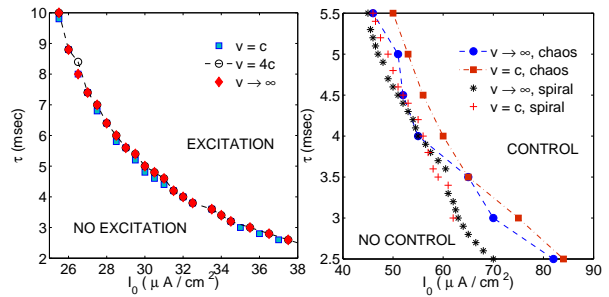}
  \caption{Strength-duration curves for LR1 model (in a two-dimensional domain
  with $L=400$)
  when control stimulation is applied on (left) quiescent homogeneous medium
  and (right) medium with existing excitation activity, either a single
  spiral or spatiotemporal chaos. In both figures, control is applied
  over a grid of points which are spaced apart by $d = 10$. Different
  curves correspond to different control signal velocities $v$,
  relative to $c$, the excitation wavefront velocity
  in the medium. For values of $I_0$ and $\tau$ above the curves, the 
  external stimulation results in (left) excitation of the domain or (right)
  control of existing activity.}
\label{fig_parameterspace}
\end{figure}

To understand in detail the lower bound on the external stimulation
parameters $I_0$ and $\tau$, we first look at the condition for exciting
a completely homogeneous medium in the resting state. The stimulation
at each point must exceed the local threshold in order to generate an
action potential. This could be achieved either directly through an
external current $I_{ext}$ or indirectly through diffusion from a
neighboring excited region. Fig.~\ref{fig_parameterspace}~(left) shows 
the result of applying control signals with different $I_0$ and $\tau$
at points which are spaced a distance $d$ apart. The resulting 
{\em strength-duration curve} \cite{Gedd70,Gold97} 
indicates that the response of the system is not sensitively dependent
on the propagation velocity $v$ of the 
control signal along the grid.
As $d$ decreases, excitation
is possible at lower values of $I_0$ and $\tau$, the minimum being for
the case when all points are subject to direct external stimulation 
($d \rightarrow 0$). This is because the entire applied 
current $I_0$ at any point is used to raise its state above the threshold,
no part being lost to neighboring regions through diffusion.

For systems with existing activity, such as self-sustaining 
spiral waves or spatiotemporal chaos, the regions in the 
relative recovery period can be excited by
stimuli larger than that needed for a fully recovered medium.
Hence, the strength-duration curve for control of such a system will
shift towards higher values of $I_0$ and $\tau$ 
(Fig.~\ref{fig_parameterspace}, right). We observe that the minimum external
stimulus required for control does not vary significantly if the medium
is undergoing fully developed spatiotemporal chaos as opposed to having a
single spiral wave. Further, the velocity of the control signal along the
array is not critical to the success in eliminating existing activity,
provided $v$ is not significantly smaller than $c$. Note that, if $v$ 
is sufficiently small or $d$ increases beyond a critical value, the control 
fails, as the effect of the control signal is confined to the region
immediately surrounding the stimulated point. 

\begin{figure}
\includegraphics[width=0.95\linewidth, clip=]{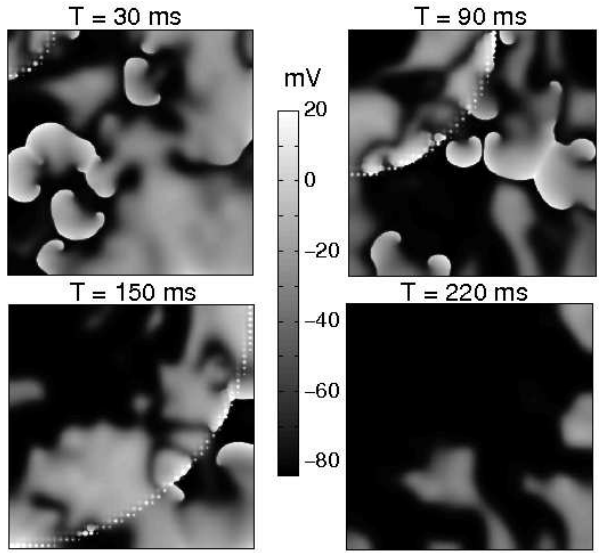}
\caption{Pseudo-gray-scale plots of the transmembrane potential $V$ for
the two-dimensional LR1 model ($L = 400$) showing the elimination of all
chaotic activity within 300 ms after initiation of control. A single wave
of control stimulus ($I_0 = 75$ $\mu$A/cm$^2$, $\tau = 3$ ms) begins at 
the top left corner ($T = 0$) and travels
across the domain with velocity $v=c$. 
This results in stimulating the region around the
control points (spaced apart by $d=10$) in a sequential manner, creating 
a stimulated wavefront
seen as an arc consisting of excited points in the panels above.
}
\label{LRcontrol}
\end{figure}
\begin{figure}
\begin{center}
  \includegraphics[width=0.3\linewidth, clip=]{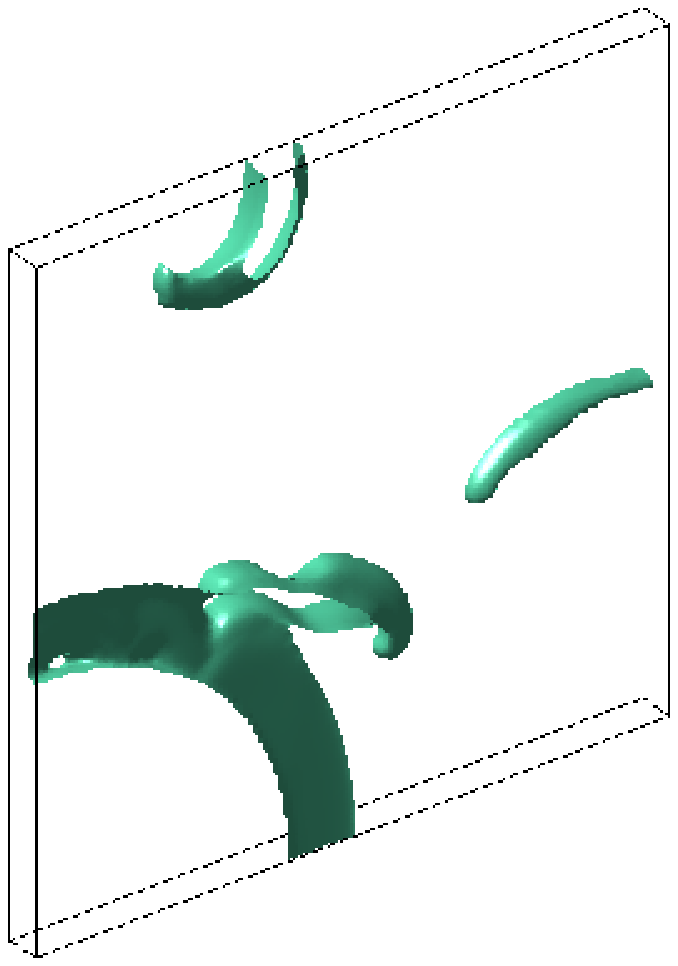}
  \includegraphics[width=0.3\linewidth, clip=]{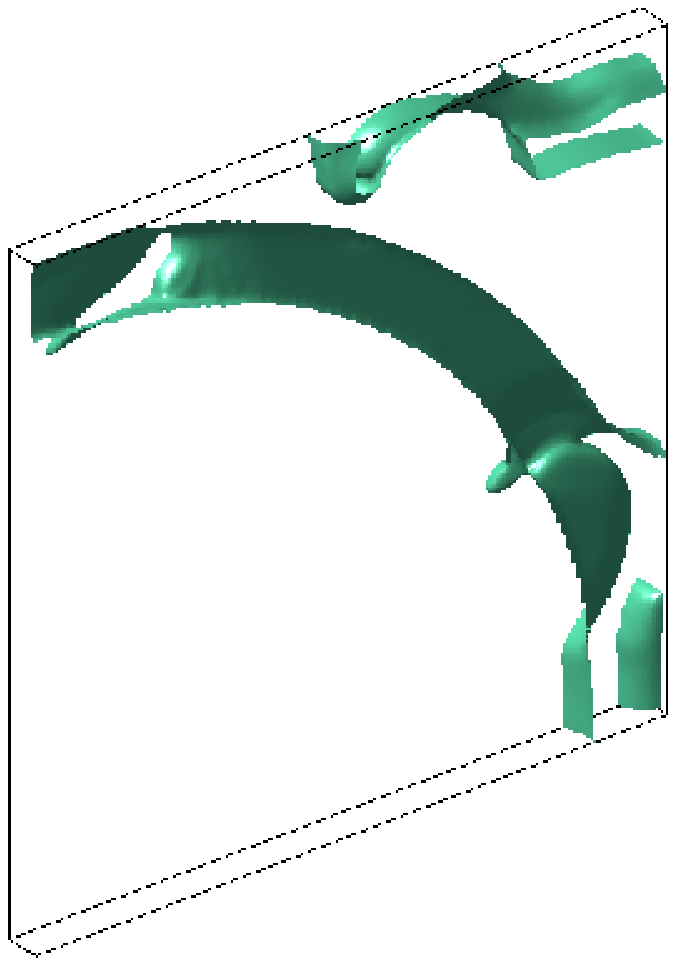}
  \includegraphics[width=0.3\linewidth, clip=]{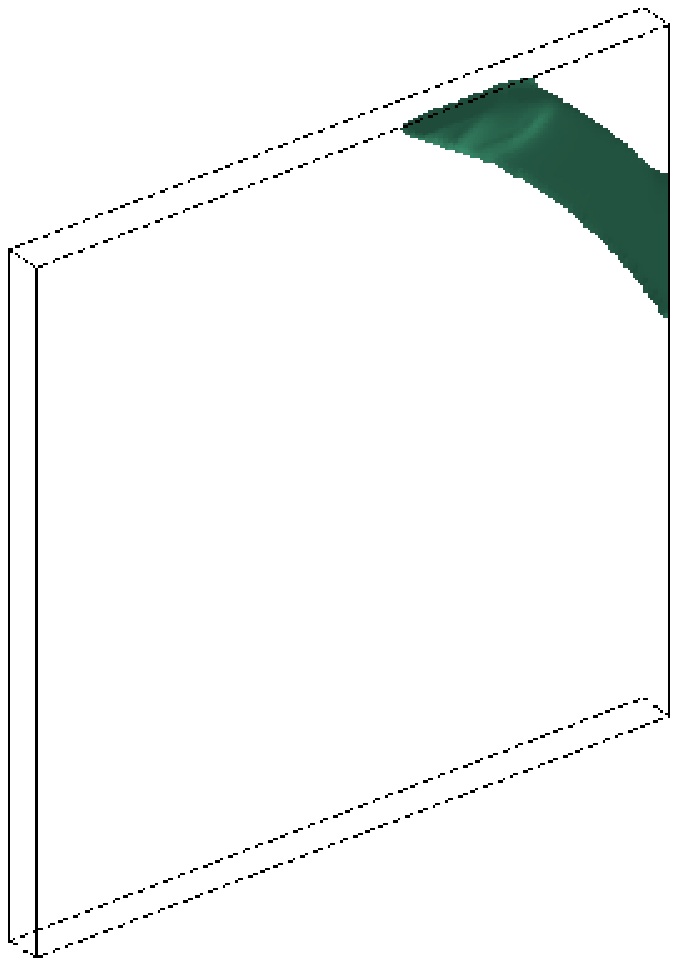}
  \end{center}
  \caption{The effect of applying control on a single surface of a 
  3-dimensional domain using the PV model ($L$ = 128 and $L_z = 8$). 
  Two waves of control signal ($I_0 = 10$, $\tau = 16.5$ ms, $v = c$)
  are applied 165 ms apart resulting in termination of spatiotemporal chaos.
  The interval between control points, $d \rightarrow 0$.
  The panels show isosurface plots at $T = 242$ ms (left), 
  308 ms (center) and 341 ms (right).}
\label{fig_3D}
\end{figure}

Fig.~\ref{LRcontrol} shows the successful termination of spatiotemporal
chaos in the LR1 model using a control signal that travels across
the 2-dimensional domain while exciting points in the medium that are
spaced apart by $d = 10$.
We verified that the control scheme is not model dependent by using
it to eliminate spatiotemporal chaotic activity in the PV model.
As most systems in reality have depth, it
is crucial to verify that the method is successful in controlling chaos
in a 3-dimensional domain,
even when the external stimulus is applied {\em only on one
surface}. This latter restriction follows from the fact that, in most practical 
situations it may not be possible (or desirable) to penetrate the medium
physically in order to apply control signals inside the bulk.
We confirmed that our method works in thin slices of excitable media
of size $L \times L \times L_z$ ($L_z \ll L$), when the array of control 
points is placed on one of the $L \times L$ surfaces (Fig.~\ref{fig_3D}).
Even in cases where a single control-stimulated wave across the medium
is unable to terminate all activity, we notice that it results in driving
the chaotic activity further towards the boundaries and away from the
origin of control stimulation. Thus, using multiple
waves through application of control signals at intervals which are larger 
than the recovery
period of the medium, the chaos in the bulk of 3-dimensional
systems is successfully terminated. 
%
%
%

\begin{figure}
\includegraphics[width=0.95\linewidth,clip]{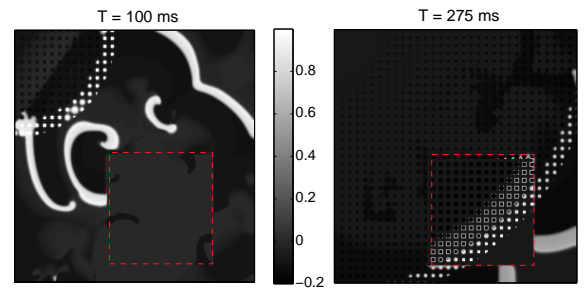}
\caption{Pseudo-gray-scale plots of the transmembrane potential $V$ for the
two-dimensional PV model ($L = 256$) showing chaos control
in the presence of a conduction
inhomogeneity (of size $110 \times 110$, indicated by the broken lines).
Inside this region, $D_{inhomogen} = 0.01 D$, the diffusion constant of
the rest of the medium. A single pulse of control stimulus ($I_0 = 8$, 
$\tau = 17$ ms, $v = c$) is applied over an array of control points spaced 
apart by $d = 8$, resulting in termination of all activity by $T = 350$ ms.}
\label{Inhomogeneity}
\end{figure}

We have checked that small distortions in the regular array of control
points does not result in the failure of our method. Similarly, starting
the control signal at different points of origin (and indeed, using
a planar wave rather than a curved wave) does not affect the efficacy
of the scheme. 
Further, our method is robust in the presence of conduction inhomogeneities 
(such as inexcitable obstacles) that tend to destabilise local control schemes.
Fig.~\ref{Inhomogeneity}
shows successful control of chaos when the medium contains a large
region of slow conduction, i.e., an extremely small value of $D$
compared to the rest of the medium.

In this paper, we have presented a 
novel control scheme involving external stimulation applied over
an array of points, that is successful in terminating spatiotemporal chaos
in both simplified as well as realistic models of biological excitable media.
The control signal amplitude is varied both spatially and temporally,
such that it appears as a propagating wave along the array of control 
points. This results in a stimulated wavefront in the excitable medium, that,
depending on the propagation velocity of the control signal and 
the space interval between control points, eliminates all existing activity.
Our method requires very low-amplitude control currents applied
for short durations at a finite number of points, each point being
stimulated once (or at most, twice) in most situations. 
Further, it is successful in terminating chaos in the bulk
of a three-dimensional medium even when applied only on one surface.
The use of significantly lower number of control points than that necessary
for global control methods, makes the proposed scheme more suitable for 
practical implementation.

We thank the IMSc Complex Systems Project and IFCPAR (Project 3404-4) 
for financial support.

\end{document}